\documentclass[]{interact}

\usepackage{float}
\usepackage{breqn}
\usepackage{epstopdf}% To incorporate .eps illustrations using PDFLaTeX, etc.
\usepackage[caption=false, skip = 0 pt]{subfig}% Support for small, `sub' figures and tables

\usepackage[para]{threeparttable}
\usepackage[labelfont=bf]{caption}
\usepackage{adjustbox}
\usepackage{multirow}
\usepackage[numbers,sort&compress]{natbib}% Citation support using natbib.sty
\bibpunct[, ]{[}{]}{,}{n}{,}{,}% Citation support using natbib.sty
\renewcommand\bibfont{\fontsize{10}{12}\selectfont}% Bibliography support using natbib.sty
\makeatletter% @ becomes a letter
\def\NAT@def@citea{\def\@citea{\NAT@separator}}% Suppress spaces between citations using natbib.sty
\makeatother% @ becomes a symbol again

\theoremstyle{plain}% Theorem-like structures provided by amsthm.sty
\newtheorem{theorem}{Theorem}[section]

\theoremstyle{definition}
\newtheorem{definition}[theorem]{Definition}

\theoremstyle{remark}

\begin{document}
\setlength{\abovedisplayskip}{5pt}
\setlength{\belowdisplayskip}{5pt}
%\articletype{ARTICLE TEMPLATE}% Specify the article type or omit as appropriate

\title{The AL-Gaussian Distribution as the Descriptive Model for the Internal Proactive Inhibition in the Standard Stop Signal Task}
\author{
\name{Mohsen Soltanifar\textsuperscript{a}\thanks{CONTACT Mohsen Soltanifar. Email: mohsen.soltanifar@mail.utoronto.ca}, Michael Escobar\textsuperscript{b}, Annie Dupuis\textsuperscript{c}, Andre Chevrier\textsuperscript{d} and  Russell Schachar\textsuperscript{e}}
\affil{\textsuperscript{a,b,c} Biostatistics Division, Dalla Lana School of Public Health, University of Toronto, Toronto, M5T 3M7, ON, Canada ;  \textsuperscript{a,c,d,e} Department of Psychiatry,  The Hospital for Sick Children, 555 University Avenue, Toronto, M5G 1X8, ON, Canada }
}

\maketitle

\begin{abstract}
Measurements of response inhibition components of reactive inhibition and proactive inhibition within the stop-signal paradigm have been of particular interest to researchers since the 1980s. While frequentist nonparametric and Bayesian parametric methods have been proposed to precisely estimate the entire distribution of reactive inhibition, quantified by stop signal reaction times(SSRT), there is no method yet in the stop-signal task literature to precisely estimate the entire distribution of proactive inhibition. We introduce an Asymmetric Laplace Gaussian (ALG) model to describe the distribution of proactive inhibition. The proposed method is based on two assumptions of independent trial type(go/stop) reaction times and Ex-Gaussian (ExG) models. Results indicated that the four parametric, ALG model uniquely describes the proactive inhibition distribution and its key shape features; and, its hazard function is monotonically increasing, as are its three parametric ExG components. In conclusion, both response inhibition components can be uniquely modeled via variations of the four parametric ALG model described with their associated similar distributional features.
\end{abstract}


\section{Introduction}

Response inhibition refers to one’s ability to stop responses or impulses that just became inappropriate or unwanted within continually changing environments,\cite{I1}. This process's importance lies in one’s being in continually changing conditions, which require new, updated courses of action,\cite{I2}. Some instances of response inhibition in daily life include braking while driving a vehicle into an intersection in reaction to a sudden traffic change, changing direction during a tennis game, and resisting an extra piece of pizza at a birthday party. Two paradigms have been proposed to study the lab setting's response inhibition, \cite{I3}: the stop-signal task and the Go/No-go task. In the standard stop-signal task, as used in this study, the task consists of a two-choice, response time task called the “go task” and the “stop task.” The go task is the primary task in which the participants are asked to correctly press a right or left button, in response to stimulus presentation, an “X” or “O” on the computer screen. The stop task is the occasional, secondary task in which (with a probability of stop signal $p_{ss}$ ) the participants are presented with a stop signal alarm after a temporal delay; Participants are instructed to withhold their responses the ongoing go task. Successful response inhibition occurs when participants successfully withhold their response to the “X” or “O” on the screen in the stop task (Figure 1). \par

\begin{figure}[H]
\begin{center}
\includegraphics[totalheight=9 cm, width=9 cm]{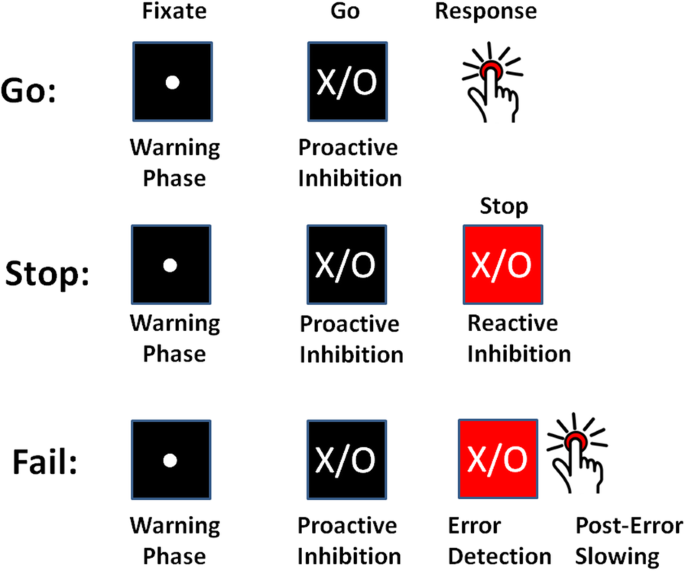}
\caption{The standard stop signal task with two inhibition components: proactive inhibition, reactive inhibition \cite{I4}}
\end{center}
\end{figure}
 
Response inhibition has two distinctive temporal-dynamic components: reactive inhibition and proactive inhibition. Both components have been utilized within the standard stop-signal task, or its varieties, to discriminate different clinical groups, \cite{I5,I6}. We refer to reactive inhibition as the outright inhibition triggered by an external cause, while proactive inhibition is restraint of actions in preparation for stopping by external conditions, \cite{I7}. Each of these inhibition components have been quantified in distinctive methods as constant point estimate  or distribution in the stop-signal task (SST) literature\cite{I1,I8,I9,I10,I11,I12,I13,I14,I15,I16,I17,I18,I19,I20,I21,I22,I23}.  \par
Reactive inhibition has been quantified as Stop Signal Reaction Times (SSRT) in the SST literature from both point estimation and distributional perspectives. Primary point estimations of the reactive inhibition up to now include the Crude SSRT, the Logan 1994 SSRT, \cite{I1}, the Weighted SSRT, the Mixture SSRT, \cite{I8}, and the time series-based SSRT,\cite{I9}. On the other hand, primary distributional estimations of the reactive inhibition include the Colonius’s nonparametric method,\cite{I10}, the Bayesian Parametric Approach (BPA), \cite{I11,I12} — with two subtype methods: the Individual BPA (IBPA) and the Hierarchical BPA(HBPA) — and the mixture method, \cite{I13}. In the mixture method, the entire SST data is partitioned to type A SST data, trials following a go trial, and type B SST data, trials following a stop trial; The trial type weights are defined as $W_A^c=$proportion of type A stops within all stops, $W_B^c=1-W_A^c.$ Now, for  $W_A\sim Bernoulli(W_A^c ), W_B\sim Bernoulli(W_B^c ),$ and fitted cluster type $SSRT_A, SSRT_B$ distributions by the former two methods, the mixture index for the reactive inhibition is defined as:\par 
\begin{eqnarray}
\label{eq1}
SSRT_{Mixture} =^d W_{A}\times SSRT_{A}+W_{B}\times SSRT_{B}.
\end{eqnarray}
Note that for the case of parametric mixture SSRT, the cluster type components  $SSRT_A,$  and  $SSRT_B$  may take a variety of proposed reaction time (RT) models, such as Ex-Gaussian (ExG), Ex-Wald, Wald, Gamma, Weibull, and lognormal, \cite{I14,I15,I16}. However, given the ExG model's practical advantages to others, it is widely considered the parametric model for the reactive inhibition,\cite{I11,I12}. \par 
Proactive inhibition has been quantified in the SST literature merely as point estimation from different perspectives. The first such estimation is defined through a dynamic Bayesian model,\cite{I17}. Here, using the mixture assumption for the predictive distribution for the probability of the kth trial being a stop task $``p(Stop),”$ the reactive inhibition is defined as the following positive Pearson correlation called the "Sequential Effect(SE)":\par 
\begin{eqnarray}
\label{eq2}
SE &=& Corr(p(Stop),GORT).
\end{eqnarray}
The second estimation is based on the variation of differences of reaction times in a go trial (GORTs) in the associated arms of the modified standard stop signal task paradigm,\cite{I18,I19,I20,I21,I22,I23}. Generally, for two given probabilities of stop signals in two arms of the modified SST where $0 \ll p_{ss}^{(1)}<p_{ss}^{(2)}\ll 1,$ the arm type proactive inhibition $\Delta_{AT} GORT$ is defined by as:\par 
\begin{eqnarray}
\label{eq3}
\Delta_{AT} GORT&=&mean(GORT_{(p_{ss}^{(2)} )} )-mean(GORT_{(p_{ss}^{(1)} )} ). 
\end{eqnarray}

The last type of the point estimation of proactive inhibition is based on the differences of GORTs in the trial type clusters of the standard SST paradigm \cite{I8,I9}(Figure.2.). Here, for the fixed stop signal probability (e.g.$p_{ss}=0.25$) and the type A GORTA and type B GORTB, the trial type proactive inhibition $\Delta_{TT} GORT$ is defined as:\par 
\begin{eqnarray}
\label{eq4}
\Delta_{TT} GORT&=&mean(GORT_B )-mean(GORT_A ). 
\end{eqnarray}

\begin{figure}[H]
\begin{center}
\includegraphics[totalheight=9 cm, width=14 cm]{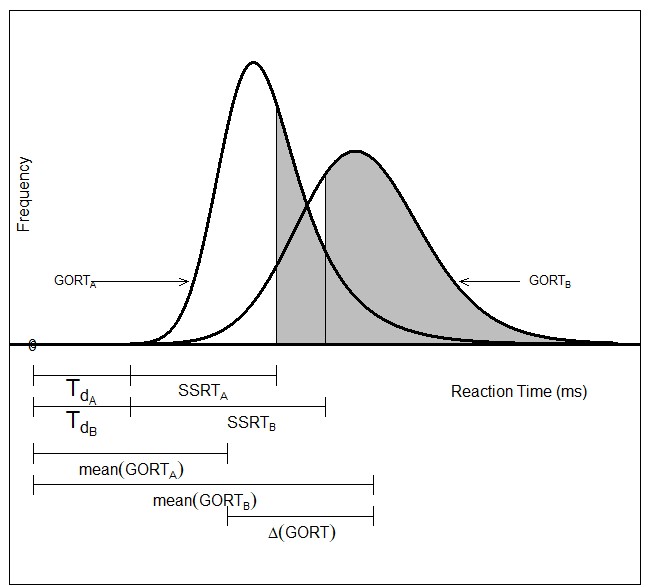}
\caption{Trial type point estimation of proactive inhibition in the standard stop signal task}
\end{center}
\end{figure}

 However, little information is available in the SST literature on the entire distribution of proactive inhibition and its key features. This subject is significant as measures of central tendencies, such as mean or median, are insufficient — and even unnecessary — to compare mostly skewed response inhibition distributions,\cite{I24}. Besides, masking prominent features of the proactive inhibitions by using central tendency measures may result in incorrect conclusions about their make-up. As an example, two different clinical groups may have the same mean of proactive inhibition, but the shape of their distributions may differ: one may be more positively skewed, or more leptokurtic, or possess a higher domain of variance. The methods mentioned earlier of the estimation of proactive inhibition do not allow for precise estimation and description of the appropriate models for the entire set of proactive inhibition distributions. \par 
This study proposes a four parametric, Asymmetric Laplace Gaussian (ALG) model for the entire proactive inhibition, given the assumption of independent trial type (go/stop) GORTs within the standard stop-signal task. The study outline is as follows. First, as in \cite{I8,I9} the overall SST data for each participant is partitioned to type A SST data and type B SST data. Second, using the Individual Bayesian Parametric Approach (IBPA) method \cite{I11,I12}, the fitted ExG GORT mean posterior parameters are calculated for the cluster type SST data.  Then, the distribution of proactive response inhibition is introduced as the difference of two independent fitted GORT ExG models, and it is shown that it has a four-parametric ALG distribution. Third, the descriptive statistics, shape statistics, and vital distributional properties of the ALG model, such as component decomposition, shape and tail behavior, and monotonic hazard function, are discussed. Finally, an empirical example is presented to manifest the above-presented results. Table~\ref{tab1} presents the summary of estimations methods presented in the literature, including this study given the type of inhibition component. 

\begin{table}[H]
\centering % centering table
\caption{Summary of Estimation Methods of Inhibition Components} %title of the table
\label{tab1}
\resizebox{\textwidth}{!}{
\begin{threeparttable}[b]
\begin{tabular}{llll} % creating four columns
\hline
            Estimation        & Inhibition Component & \\
\hline
                              & Reactive & Proactive \\ 
\hline
            Constant Index& SSRT     &  $\Delta GORT$,$SE$ \\
            Examples  & $SSRT_{Crude},SSRT_{Mixture},SSRT_{Weighted}$ & $\Delta_{AT}GORT,\Delta_{TT} GORT$ \\
                     & $SSRT_{Logan1994},SSRT_{SS.Logan1994}$&  \\
            Distribution Index& SSRT  &  $\Delta GORT$ \\
            Examples  &  ExG,LN,Wald            & ALG\\ 
                      &  Ex-Wald, Gamma         &    \\                            
\hline
\end{tabular}
\end{threeparttable}
}
\end{table}

\section{Materials \& Methods}

\subsection{Data}
The study data have been previously described \cite{M1,I8,I9,I13}. Data were collected at the Ontario Science Center in Toronto, Canada, in 2009-2010. Included were 16,099 participants aged 6 to 17. The participants’ parents provided the required ethical consent for the SST experiment. Each participant completed the SST task, including four blocks of 24 trials with a total of 96 trials, including random 25\% stop trials (24 stops) and 75\% go trials(72 goes). The SST tracking algorithm was designed so that at the end of trials, each participant achieves a 50\% probability of successful inhibition. \par 

\subsection{Participants}
The study participants are the same as those described in \cite{I9,I13}. Included here is a unique subsample of 44 participants with a mean age of 12.1 years, with 96 SST trials for each, and an almost balanced number of trial type stop trials (10-14 type B stop trials vs. 14-10 type A stop trials, respectively). This almost balanced number of trial type stop trials yields 30-42 type B go trials vs. 42-30 type A stop trials, respectively.\par 

\subsection{The SST Clusters}

The study Stop Signal Task clusters have been described before in \cite{I8,I9,I13}. Each participant’s SST data was partitioned to type A SST data, where a go trial preceded all trials, and type B SST data, all trials preceded by a stop trial. Hence, each participant has three types of SST data clusters: Type-A SST cluster (i.e., 56 trials), Type B SST cluster(i.e., 40 trials), and Type-S SST cluster (all 96 trials). Then, using IBPA, the parameters of the corresponding Ex-Gaussian (ExG) GORT’s parameters (i.e. 
$\theta_A=(\mu_A,\sigma_A,\tau_A),\theta_B=(\mu_B,\sigma_B,\tau_B),\theta_S=(\mu_S,\sigma_S,\tau_S) $ were computed as described in the upcoming results section 3.2. \par 

\subsection{Preliminaries on Component Distributions}
The first two definitions provide the critical features of the components of our upcoming calculations, namely Ex-Gaussian distribution (ExG), \cite{M2}, and the Asymmetric Laplace distribution (AL), \cite{M3}.\par   
\begin{definition}
\label{def2.1}
A random variable has an Ex-Gaussian (ExG) distribution with parameters $(\mu,\sigma,\tau)$ whenever it is considered as the sum of an independent normal random variable with parameters $(\mu,\sigma^2)$ and an exponential random variable with parameter $\tau$: 
\begin{eqnarray}
\label{eq5}
ExG(\mu,\sigma,\tau) &=^d& N(\mu,\sigma^2)\oplus Exp(\tau).
\end{eqnarray}
\end{definition}
The density, the moment generating function, the nth cumulant $(n\geq 1),$ the variance, the skewness and the kurtosis of the ExG distribution are given by:\par 

\begin{eqnarray}
\label{eq6}
PDF && f_{ExG} (t|\mu,\sigma,\tau)=\frac{1}{\tau} exp(\frac{\mu-t}{\tau}+\frac{\sigma^2}{2\tau^2})*\Phi(\frac{\mu-t}{\sigma}-\frac{\sigma}{\tau}): \sigma,\tau>0, t\in\mathbb{R}, \nonumber\\
MGF&&m_{ExG}(t)=(1-t\tau)^{-1} exp(\mu.t+\frac{\sigma^2}{2}t^2): t<\tau^{-1}, \nonumber\\
n^{th} Cumulant&& \kappa_{n}^{ExG}=(n-1)!\tau^{n}+1_{n=1}(n)\mu+1_{n=2}(n)\sigma^2: 1\leq n, \nonumber\\
Mean &&E(ExG)= \mu+\tau, \nonumber\\
Variance &&Var(ExG)=\sigma^2+\tau^2, \nonumber\\
Skewness &&\gamma_{ExG}=2(1+\sigma^2 \tau^{-2} )^{-3/2}, \nonumber\\
Kurtosis &&\kappa_{ExG}=3 \frac{(1+2\sigma^{-2} \tau^2+3\sigma^{-4} \tau^4)}{(1+\sigma^{-2} \tau^2 )^2 }. 
\end{eqnarray}

\begin{definition}
\label{def2.2}
A random variable has an Asymmetric Laplace (AL) distribution with parameters $(\alpha_1,\alpha_2)$ whenever it is considered as the difference of two independent exponential random variables with parameters $\alpha_2$,  and $\alpha_1$, respectively:
\begin{eqnarray}
\label{eq7}
AL(\alpha_1,\alpha_2 ) &=^d& Exp(\alpha_2 ) \ominus Exp(\alpha_1 ).
\end{eqnarray}
\end{definition}

The density, the moment generating function, the nth cumulant $(n\geq1),$ the variance, the skewness and the kurtosis of the AL distribution are given by:\par

\begin{eqnarray}
\label{eq8}
PDF&& f_{AL}(t|\alpha_1,\alpha_2)= \frac{ exp(\frac{t}{\alpha_1} ) 1_{(-\infty,0)} (t)+ exp(\frac{-t}{\alpha_2} ) 1_{[0,\infty)}  (t)}{\alpha_1+\alpha_2}\ t\in\mathbb{R}, \nonumber\\
MGF&& m_{AL}(t)= (1+(\alpha_1-\alpha_2)t-\alpha_1\alpha_2.t^2)^{-1}\  -\alpha_1^{-1}<t<\alpha_2^{-1},\nonumber\\
n^{th} Cumulant&& \kappa_{n}^{AL}=(n-1)!((-\alpha_1)^n+(\alpha_2)^n)\ 1\leq n, \nonumber\\
Mean && E(AL)= -(\alpha_1-\alpha_2), \nonumber\\
Variance && Var(AL)=\alpha_1^2+\alpha_2^2, \nonumber\\
Skewness && \gamma_{AL}=-2(\alpha_1^3-\alpha_2^3 )\times (\alpha_1^2+\alpha_2^2 )^{-3/2},\nonumber\\
Kurtosis && \kappa_{AL}=3(3\alpha_1^4+2\alpha_1^2 \alpha_2^2+3\alpha_2^4)\times(\alpha_1^2+\alpha_2^2 )^{-2}.
\end{eqnarray}
The convolution of two independent, AL random variables and Gaussian random variables, called ALG or Normal-Laplace (NL) random variables, has been of special attention in the literature \cite{M4,M5}. \par 

\begin{definition}
\label{def2.3}
A random variable has Asymmetric Laplace-Gaussian (ALG) distribution with parameters $(\alpha_1,\alpha_2,\mu,\sigma)$ whenever it is considered as the sum of two independent, Asymmetric Laplace random variables with parameters  $(\alpha_1,\alpha_2)$, and a Normal random variable with parameters $(\mu,\sigma^2)$, respectively:
\begin{eqnarray}
\label{eq9}
ALG(\alpha_1,\alpha_2,\mu,\sigma) &=^d& AL(\alpha_1,\alpha_2 )\oplus N(\mu,\sigma^2) .
\end{eqnarray}
\end{definition}
Note that since  $ AL(0^{+},\alpha_2 ) =^{d} Exp(\alpha_2 ),$ it follows that $ALG(0^{+},\alpha_2,\mu,\sigma) =^{d} ExG(\mu,\sigma,\alpha_2)$. Consequently, the ExG model can be considered a special degenerate ALG model. Next, the following key Theorems equip us to propose the ALG distribution as the model for the proactive inhibition and compute the key descriptive and shape statistics of the ALG distribution in terms of its Laplacian and Gaussian components,\cite{M6}.\par 
\begin{theorem}
\label{thm2.4}
Let $X,Y$ be two independent, real-valued random variables with finite, moment generating function $m_X,m_Y,$ and cumulant functions $\kappa,\kappa,$ respectively. Then, for some $s_0>0:$
\begin{eqnarray}
\label{eq1011}
m_{X+Y} (t)=m_X (t) m_Y (t):\ \     (-s_0<t<s_0 ),\\
\kappa_{X+Y} (t)=\kappa_X (t)+\kappa (t):\ \ \   (-s_0<t<s_0 ).
\end{eqnarray}
\end{theorem}

\begin{theorem}
\label{thm2.5}
Let $X,Y$ be two real-valued random variables with  finite moment generating functions $m_X,m_Y,$ respectively. Assume for some $s_0>0:\ m_X (t)=m_Y (t)\ (-s_0<t<s_0).$ Then, $X,Y$ have the same distribution.   
\end{theorem}

Finally, the last two theorems enable us to describe the behavior of the hazard function of the ALG model for the proactive inhibition, \cite{M7}. \par 

\begin{theorem}
\label{thm2.6}
Let $X$ be a real-valued random variable with differentiable PDF $f_X$ and CDF $F_X$ such that $f_X (t)\rightarrow 0,F_X (t)\rightarrow 1 \ as\  t\rightarrow \infty,$ and $-ln (f_X (t)) $ is convex(concave). Then, the hazard function $h_X$ is increasing(decreasing).
\end{theorem}

\begin{theorem}
\label{thm2.7}
Let $X,Y$ be two independent, real-valued random variables with (strictly) increasing hazard functions $h_X,h_Y,$ respectively. Then, the hazard function of their sum, $h_{X+Y}$ is (strictly) increasing as well. 
\end{theorem}

\subsection{Proactive Inhibition Index }

Proactive inhibition was operationalized based on the standard stop signal task's internal perspective,\cite{I9}. Here, for a given fixed stop signal probability (e.g., 0.25), type A GORT of $GORT_A$ (GORT for a trial after a go trial), and type B GORT of $GORT_B$ (GORT for a trial after a stop trial), the internal proactive inhibition is defined as:\par 
\begin{eqnarray}
\label{eq12}
\Delta GORT &=^d& GORT_B-GORT_A.
\end{eqnarray}

Note that there are two mathematical perspectives for the proactive inhibition: First, a model with two ExG components; Second, a model with Asymmetric Laplace (AL) and Gaussian components. Henceforward, it is understood within the given context which perspective is being discussed.\par 

\subsection{Statistical Analysis}

The statistical ALG model to describe the proactive inhibition distribution was presented using moment generating functions, \cite{M5}. Next, the ALG model's descriptive and shape statistics were computed in terms of parameters of the cluster type ExG components,\cite{M6}. Finally, its hazard function behavior was theoretically inferred using its components’ parameters \cite{M7}.\par 

The ExG components of the presented statistical model were estimated using the IBPA method,\cite{I11,I12}. As in \cite{I13}, each participant had three IBPA associated ExG parametric estimations $\theta_A=(\mu_A,\sigma_A,\tau_A ),  \theta_B=(\mu_B,\sigma_B,\tau_B ),$ and $\theta_S=(\mu_S,\sigma_S,\tau_S ),$ associated to type-A cluster SST data, type-B cluster SST data, and the entire SST data, respectively. These parameters were estimated as the posterior means of the associated following IBPA procedure with three chains, 5,000 burn-ins within 20,000 simulations in Bayesian Ex-Gaussian Estimation of Stop-Signal RT distributions (BEESTS) 2.0 software,\cite{I12}:\par 

\begin{tabular}{ll}
 Data & Individual Priors\\
$GORT \sim ExG(\mu_{go},\sigma_{go},\tau_{go} )$ &\\
$SRRT\sim ExG(\mu_{go},\sigma_{go},\tau_{go},\mu_{stop},\sigma_{stop},\tau_{stop} ,SSD) I^{+}_{ [1,1000]}$ & $\mu_{go},\sigma_{go},\tau_{go}\sim U[10,2000]$\\
$SSRT\sim ExG(\mu_{go},\sigma_{go},\tau_{go},\mu_{stop},\sigma_{stop},\tau_{stop} ,SSD) I^{+}_{ [1,1000]}$ & $\mu_{go},\sigma_{go},\tau_{go}\sim U[10,2000]$
\end{tabular}

Two sets of comparisons were conducted using paired t-tests (DescTools, $R$ software version 4.0.0, \cite{M8}): First, a primary comparison between the cluster-type fitted parameter of the ExG distribution, the descriptive statistics, and the shape statistics; second, secondary comparisons between the ALG model descriptive and shape statistics and its associated cluster-type ExG components.\par 

\section{Results}
 
The results are divided into two subsections. In subsection 3.1, we explore the mathematical analysis of the proposed model for the proactive inhibition in the standard stop-signal task. This model includes a four parametric ALG for the proactive inhibition and its prominent distributional properties. In subsection 3.2, we present an empirical example of the case and discuss its various distributional features.\par 

\subsection{Mathematical Analysis}

\subsubsection{The Proactive Inhibition Distribution and its Parameters }

First of all, we propose a mathematical model for the proactive inhibition provided by the ALG: \par 

\begin{theorem}
\label{thm3.1}
\textbf{(The Main Result)}. The four parametric $ALG(\tau_A,\tau_B,\mu_B-\mu_A,(\sigma_B^2+\sigma_A^2 )^{1/2})$ presents a model for the Internal Proactive Inhibition Index $\Delta GORT$ with trial type-related parameters $(\mu_A,\sigma_A,\tau_A,\mu_B,\sigma_B,\tau_B)$  in the Standard Stop Signal Task. 
\end{theorem}
As a corollary of Theorem 3.1, the probability density function $(f_{\Delta GORT})$ and the cumulative density function $(F_{\Delta GORT})$ of the ALG distribution for the Internal Proactive Inhibition Index  are given by \cite{M4,M5},:\par 

\begin{eqnarray}
\label{eq1314}
f_{\Delta GORT}(t)&=& \frac{1}{\tau_A+\tau_B}  \nonumber\\
&\times [& e^{(\frac{\sqrt{\sigma_B^2+\sigma_A^2}}{2\tau_B} (\frac{\sqrt{\sigma_B^2+\sigma_A^2}}{\tau_B}-2\frac{t-(\mu_B-\mu_A)}{\sqrt{\sigma_B^2+\sigma_A^2}}))} \times (1-\Phi(\frac{\sqrt{\sigma_B^2+\sigma_A^2}}{\tau_B}-\frac{t-(\mu_B-\mu_A)}{\sqrt{\sigma_B^2+\sigma_A^2}}))\nonumber\\
&+&  e^{(\frac{\sqrt{\sigma_B^2+\sigma_A^2}}{2\tau_A} (\frac{\sqrt{\sigma_B^2+\sigma_A^2}}{\tau_A}+2\frac{t-(\mu_B-\mu_A)}{\sqrt{\sigma_B^2+\sigma_A^2}}))} \times (1-\Phi(\frac{\sqrt{\sigma_B^2+\sigma_A^2}}{\tau_A}+\frac{t-(\mu_B-\mu_A)}{\sqrt{\sigma_B^2+\sigma_A^2}})) ]\nonumber\\
&& \hspace{10 cm} t\in\mathbb{R}\\
\text{and}&&\nonumber\\
F_{\Delta GORT}(t)&=& \frac{1}{\tau_A^{-1}+\tau_B^{-1}}\times [(\tau_A^{-1}+\tau_B^{-1}) \Phi(\frac{t-(\mu_B-\mu_A)}{\sqrt{\sigma_B^2+\sigma_A^2}}) \nonumber\\
&-&\tau_A^{-1} e^{(\frac{\sqrt{\sigma_B^2+\sigma_A^2}}{2\tau_B} (\frac{\sqrt{\sigma_B^2+\sigma_A^2}}{\tau_B}-2\frac{t-(\mu_B-\mu_A)}{\sqrt{\sigma_B^2+\sigma_A^2}}))} \times (1-\Phi(\frac{\sqrt{\sigma_B^2+\sigma_A^2}}{\tau_B}-\frac{t-(\mu_B-\mu_A)}{\sqrt{\sigma_B^2+\sigma_A^2}}))\nonumber\\
&+&\tau_B^{-1} e^{(\frac{\sqrt{\sigma_B^2+\sigma_A^2}}{2\tau_A} (\frac{\sqrt{\sigma_B^2+\sigma_A^2}}{\tau_A}+2\frac{t-(\mu_B-\mu_A)}{\sqrt{\sigma_B^2+\sigma_A^2}}))} \times (1-\Phi(\frac{\sqrt{\sigma_B^2+\sigma_A^2}}{\tau_A}+\frac{t-(\mu_B-\mu_A)}{\sqrt{\sigma_B^2+\sigma_A^2}})) ]\nonumber\\
&& \hspace{10 cm} t\in\mathbb{R},
\end{eqnarray}
respectively. Here $\Phi$ denotes the standard normal cumulative distribution function.\newline 
Next, given trial type parameters, we estimate the descriptive and shape statistics for the proposed ALG model of the proactive inhibition: \par 

\begin{theorem}
\label{thm3.2}
The descriptive statistics and the shape statistics of the Proactive Inhibition ALG distribution with trial type-related parameters $(\mu_A,\sigma_A,\tau_A,\mu_B,\sigma_B,\tau_B)$  in the standard stop signal task are given by:\par 
\end{theorem}

\begin{eqnarray}
\label{eq15}
n^{th} Cumulant&& \kappa_{n}^{ALG}=(n-1)!((-\tau_A )^n+\tau_B^n )\nonumber\\
&&\hspace{1 cm}+ 1_{(n=1)} (n)(\mu_B-\mu_A )+1_{(n=2)} (n)  ( \sigma_B^2+\sigma_A^2):\ \        1\leq n\nonumber\\
Mean && E(ALG)=  \tau_B-\tau_A+\mu_B-\mu_A, \nonumber\\
Variance && Var(ALG)= \tau_A^2+\tau_B^2+\sigma_A^2+\sigma_B^2,\nonumber\\
Skewness && \gamma_{ALG}=\frac{2(\tau_B^3-\tau_A^3 )}{(\tau_A^2+\tau_B^2+\sigma_A^2+\sigma_B^2 )^{3/2}} , \nonumber\\
Kurtosis && \kappa_{ALG}=\frac{6(\tau_B^4+\tau_A^4 )}{(\tau_A^2+\tau_B^2+\sigma_A^2+\sigma_B^2 )^2} . 
\end{eqnarray}

\subsubsection{The Proactive Inhibition's Key ALG Distributional Properties }

In this section, we present key distributional properties for the ALG model for proactive inhibition including: (i) component decompositions in terms of trial type GORT; (ii) shape and tail behavior; and, (iii) behavior of the hazard function.\par 

\begin{theorem}
\label{thm3.3}
\textbf{(Component Decomposition).} An ALG model for proactive inhibition emerges from uncountable pairs of trial type related GORT $(GORT_{A},GORT_{B})$ distributions.
\end{theorem}

We remind the reader that Theorem3.3 presents a process to simulate a plausible four parametric ALG distribution for the proactive inhibition.\par 

\begin{theorem}
\label{thm3.4}
\textbf{(Shape and Tail Behavior).}  An ALG model for proactive inhibition has unimodal, generally asymmetric, infinite, differentiable density with extreme large values proportionate to the  $Exp(1/\tau_B)$ distribution. 
\end{theorem}

We remind the reader that contrary to the ALG model's mean for proactive inhibition, there are no closed-form formulas for the mode and the median, respectively. Similar to the ExG model for reactive inhibition with increasing hazard function, we have:\par 

\begin{theorem}
\label{thm3.5}
\textbf{(Hazard Function’s Behavior).} An ALG model for proactive inhibition has increasing hazard function. 
\end{theorem}

\subsection{The Empirical Example}

\subsubsection{The ALG Model for Proactive Inhibition}

This section presents an example of the empirical data for the theoretical results inferred in the previous section on the ALG model for proactive inhibition and its descriptive, shape, and hazard function’s key features. These results are based on the cluster type IBPA estimation of mean posterior ExG parameters $\theta=(\mu,\sigma,\tau)$ presented in Table~\ref{tableB1}(Appendix B).\par  
The ExG model for each of the two components of proactive inhibition has the following key features presented by Table~\ref{tab2}.  First, while type B $\mu,$, and $\sigma$ parameters are significantly larger than their type A counterparts, there is no difference for the $\tau$ parameter. In addition, the sample average proactive inhibition is 92.1 ms (95\% CI = (69.4,114.9)). Second, both ExG components are positively skewed and leptokurtic. Finally, there is no significant difference between their trial-type skewness and their trial-type kurtosis, respectively. Figure3a presents the trial type ExG modeled components of the ALG model. \par

\begin{table}[H]
\centering % centering table
\caption{Descriptive and paired t-test [mean ($95\%CI$)] results for parameters, descriptive and shape statistics of fitted Ex-Gaussian distribution to cluster type GORT and AL-Gaussian distribution to $\Delta GORT, (n=44).$} %title of the table
\label{tab2}
\resizebox{\textwidth}{!}{
\begin{threeparttable}[b]
\begin{tabular}{l ccccc} % creating four columns
\hline
           & & & ExG model& & ALG model\\
\hline
           & &Cluster &  & Comparison & Cluster\\
\hline           
           & & Type A & Type B & Type B vs. Type A & Type S\\
\hline
           & $\alpha_1$ &- &- &- &104.2\\
           &            &- &- &- &(90.4,117.9) \\
           & $\alpha_2$ &- &- &- &142.4 \\
           &            &- &- &- &(125.9,158.8) \\            
Parameter  & $\mu$      &478.8 &532.8&53.9*** &53.9 \\
           &            &(448.0,509.7) &(498.6,566.9) &(30.9,76.9) &(30.9,76.9) \\
           & $\sigma$   &109.9 &133.1 &23.2 &179.2 \\
           &            &(90.5,129.3) &(108.4,157.8) &(-0.1,46.4) & (151.4,206.9)\\           
           & $\tau$     &104.2 &142.4 &38.2*** &- \\
           &            &(90.4,117.9) &(125.9,158.8) &(19.6,56.8) &- \\           
\hline
           & Mean       & 583.0& 675.1&92.1*** &92.1 \\
           &            &(553.0,612.9) &(633.8,716.4) &(69.4,114.9) &(69.4,114.9) \\           
Statistics & St.D       &160.6 &202.4 &41.8*** &260.4 \\
           &            &(143.5,177.8) &(177.9,226.9) &(25.9,57.6) &(232.3,288.6) \\
           & Skewness   &0.787 &0.918 &0.131 &0.186 \\ 
           &            &(0.602,0.973) &(0.751,1.085) &(-0.113,0.375) &(0.076,0.296) \\           
           & Kurtosis   &4.966 &5.300 &0.334 &1.153 \\
           &            &(4.414,5.518) &(4.790,5.808) &(-0.397,1.064) &(0.923,1.384) \\           
\hline
\end{tabular}
\begin{tablenotes}
     \item \normalsize{Notes: ${}^{*}$p-value$<0.05$;${}^{**}$p-value$<0.005$;${}^{***}$p-value$<0.0005$.}
 \end{tablenotes}
\end{threeparttable}
}
\end{table} 

The ALG model for proactive inhibition has the following features, given cluster type parameter estimations. First, as a primary corollary of Theorem 3.2, the model is positively skewed whenever  $\tau_B>\tau_A.$  The negatively skewed and symmetric cases hold whenever the strict greater inequality $>$ is replaced with  $<$ and $= ,$ respectively. According to appendix Table ~\ref{tableB1} data, all three cases exist (case 10: positive skew; case 16: symmetric; case 11: negative skew). Figure 3b presents all the mentioned three cases.  Overall, the model is positively skewed given the results in Table~\ref{tab2}. Second, as the second corollary of Theorem 3.2, the model is leptokurtic whenever
$(2(\tau_A^4+\tau_B^4))^{1/2}-(\tau_A^2+\tau_B^2)>\sigma_A^2+\sigma_B^2.$
The platykurtic and mesokurtic cases hold whenever the strict greater inequality $>$ is replaced with $<$ and $=,$ respectively. In particular, for the case, $\tau_A\approx\tau_B,$   the model is always platykurtic. Overall, the model is platykurtic given results in Table ~\ref{tab2}.  Third, while the ALG model’s standard deviation is larger than its two ExG’s components, its skewness and kurtosis are significantly smaller. Finally, the ALG model has a strictly increasing hazard function for various skewness cases, as mentioned in Theorem3.5 and presented in Figure 3c. \par 

\begin{figure}[H]
\begin{center}
\includegraphics[totalheight=18 cm, width=14 cm]{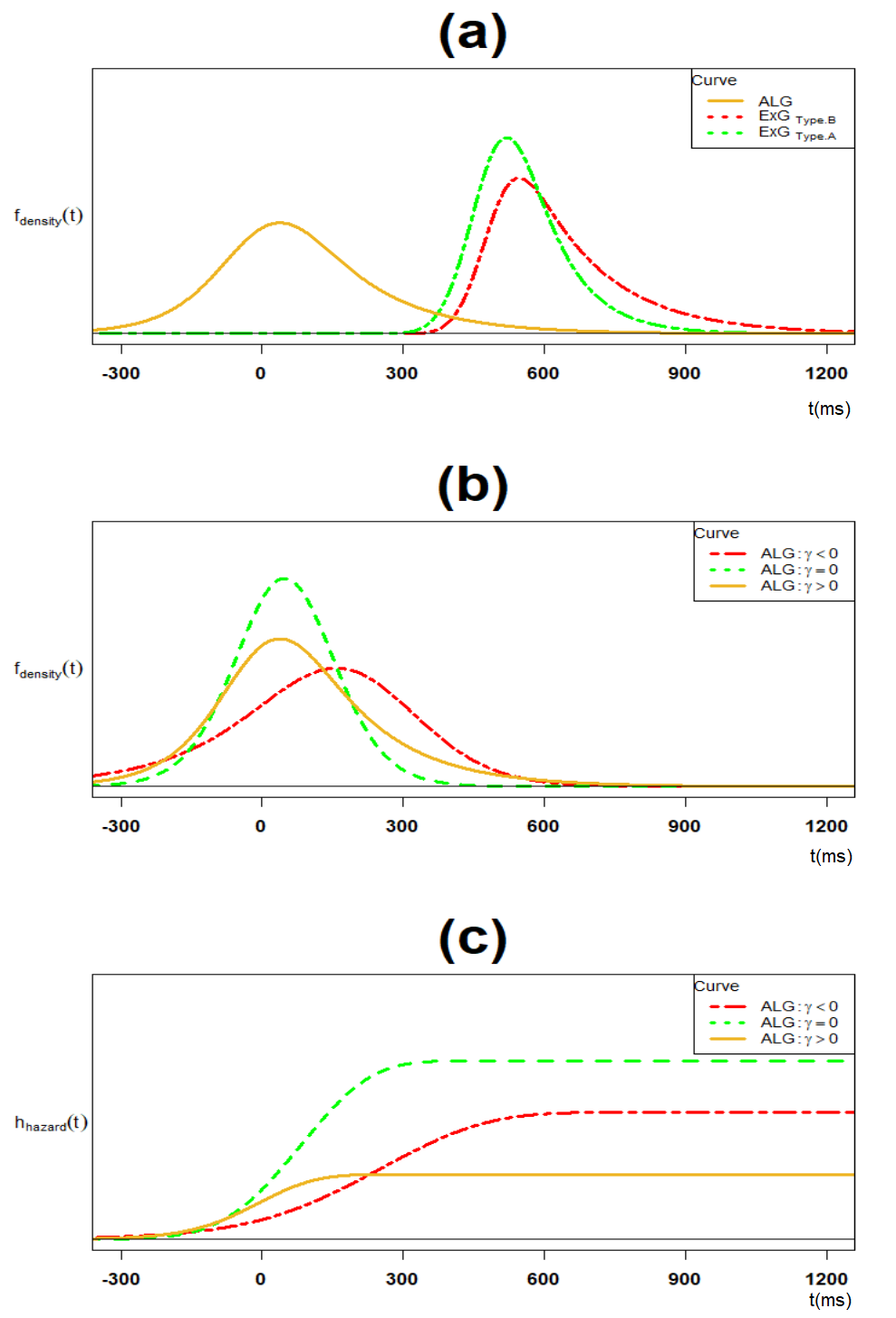}
%\vspace*{\floatsep}
\caption{The ALG density and its trial type ExG component densities; (b) The ALG density for the positively skewed, symmetric and negatively skewed cases; (c) The ALG hazard function for the positively skewed, symmetric and negatively skewed cases.}
\end{center}
\end{figure}
 
\subsubsection{Proactive Inhibition ALG Model versus Reactive Inhibition ExG Model}

This section compares the ALG distribution of the proactive inhibition and the ExG distribution of the reactive inhibition in descriptive and shapes statistics. Table~\ref{tab3} presents the corresponding statistics for both models. As it is seen, the proactive ALG inhibition distribution has a significantly lower mean, lower skewness, lower kurtosis, and higher standard deviation than reactive inhibition ExG distribution. Also, while the proactive inhibition ALG distribution is platykurtic, the corresponding reactive ExG distribution is leptokurtic. However, the two distributions are both positively skewed. Overall, the two distributions are significantly distinctive. \par

\begin{table}[H]
\centering % centering table
\caption{Comparison of proactive inhibition ALG model versus reactive inhibition ExG model in terms of descriptive and shape statistics $(n=44).$} %title of the table
\label{tab3}
\resizebox{\textwidth}{!}{
\begin{threeparttable}[b]
\begin{tabular}{l cccc} % creating four columns
\hline
Inhibition & &Reactive & Proactive & Proactive vs. Reactive\\
Index      & & $SSRT$  & $\Delta GORT$ & $\Delta GORT\ vs.\ SSRT$ \\
Model      & &  ExG    & ALG & ALG vs. ExG\\
\hline
           & Mean       & 196.8          & 92.1         & -104.6*** \\
           &            & (173.5,220.1)  & (69.4,114.9) & (-140.6,-68.7) \\           
Statistics & St.D       & 157.8          & 260.4        &102.6*** \\
           &            & (139.4,176.2)  & (232.3,288.6)&(71.8,133.6)  \\
           & Skewness   & 0.578        & 0.186        &-0.401***  \\ 
           &            & (0.500,0.674)  & (0.076,0.296)&(-0.540,-0.261)  \\           
           & Kurtosis   & 4.231          & 1.153        &-3.077***   \\
           &            & (3.998,4.465)  & (0.923,1.384)&(-3.381,-2.775) \\           
\hline
\end{tabular}
\begin{tablenotes}
     \item \normalsize{Notes: ${}^{*}$p-value$<0.05$;${}^{**}$p-value$<0.005$;${}^{***}$p-value$<0.0005$.}
 \end{tablenotes}
\end{threeparttable}
}
\end{table}

\section{Discussion}
\subsection{Present Work}
 This paper presents a four parametric model for the entire proactive inhibition distribution, the ALG model. This model is based on the two independent ExG components fitted to the trial type GORTs. Considering ExG models as a degenerate ALG model, this work indicates that a four parametric ALG model can model both response inhibition components. \par 

The proposed ALG model for proactive inhibition has several important aspects. First, the model is based on the independent assumption of GORTA and GORTB. Such speculation is warranted by the independent assignment of go trials and stop trials in the entire stop-signal task trials. Hence, via conditioning, their following associated type A go trial and type B go trials are independent, and so are their associated GORTs (i.e., GORTA and GORTB). Second, contrary to the point estimations of proactive inhibition in the form of mean \cite{I9}, or correlation \cite{I17}, it presents the entire distribution of the proactive inhibition.  Third, the ALG model for proactive inhibition is entirely distinctive from the ExG model for reactive inhibition in terms of the mean, standard deviation, and kurtosis. This result provides more evidence on the distinction between proactive inhibition and reactive inhibition as a whole distribution. Fourth, similar to the ExG model for reactive inhibition, the ALG model for proactive inhibition is skewed to the right and has a monotonically increasing hazard function. Finally, the ALG model for proactive inhibition with its associated ExG modeled components is unique because other non-ExG RT models for its components (e.g., Gamma, Weibull, Lognormal, Wald, and Ex-Wald) do not yield to any known closed-form distribution for proactive inhibition. This limitation is easily verifiable by repeating the proof of Theorem 3.1 based on the uniqueness of moment generating functions for other non-ExG RT models for the trial type GORTs. \par 

The proposed ALG model for the proactive inhibition is estimated (or simulated) in two different methods. First, one may use the Bayesian or frequentist-based methods to fit the ExG parametric models to its trial-type components and then estimate the four parametric ALG model using Theorem 3.1, as done in Section 3.2. Second, one may subtract the trial type GORTs and fit the ALG model directly to the differenced GORT data using Maximum Likelihood (ML) or Expectation-Maximization (EM) algorithms \cite{D1}.\par 

The proposed ALG model uniquely distinguishes the reactive inhibition and proactive inhibition distribution in terms of vital distributional features. Table~\ref{tab4} presents an overall comparison for proactive inhibition and reactive inhibition in terms of the ALG model(with considering ExG as its particular case): \par

\begin{table}[H]
\centering % centering table
\caption{Comparison of proactive inhibition and reactive inhibition in terms of ALG model properties.} %title of the table
\label{tab4}
\resizebox{\textwidth}{!}{
\begin{threeparttable}[b]
\begin{tabular}{l cccccccc} % creating four columns
\hline
Inhibition & Index & $\#$ Parameters & $\#$ Estimations & Mean & StD & Skewness($+$) & Kurtosis & Hazard\\      
\hline
Proactive & $\Delta GORT$ & 4 & 2 & lower & higher & lower & platykurtic & increasing\\
Reactive  & $SSRT$ & 3 & 1 & higher & lower & higher & leptokurtic & increasing\\
\hline
\end{tabular}
\end{threeparttable}
}
\end{table} 

There are some limitations in the proposed ALG model for proactive inhibition. First, since GORTA data and GORTB data are unmatched, there is no way to calculate their correlation quantitatively.  Hence, from a quantitative perspective, checking the validity of the assumption of independent GORTA and GORTB is difficult. Second, by its definition, proactive inhibition takes only non-negative values while the presented ALG model takes negative values. Third, similar to the ExG model for reactive inhibition, the ALG model for proactive inhibition has a monotonically increasing hazard function preventing it from being the best fitting model for the cases of proactive inhibition with peaked hazards. Finally, given the calculations' structure of the ALG model parameters, based on those of ExG components, its parameters' cognitive interpretations are highly dependent on its ExG components inheriting their constraints. \par

\subsection{Future Work}
Future research should replicate the proposed approach in modeling the proactive inhibition distribution in this study in other different directions. This further work may include the following perspectives: First, one model should consider peaked hazard functions for the ALG model components to address RT data with such features. Second, there is a need to interpret the proposed ALG distribution parameters in terms of inhibition mechanisms in the brain and vise versa. Third, there is a lack of investigation comparing the proactive inhibition distribution in terms of the usual stochastic order, the descriptive and shape statistics across a spectrum of clinical groups such as ADHD, OCD, schizophrenia, and drug users. Finally, similar investigations on comparing the proactive inhibition distribution and its above key statistics are plausible in terms of the participants' age.\par 

\subsection{Conclusion}
In conclusion, the ALG model provides a practical description of the proactive inhibition distribution that takes full advantage of its ExG components fitted for the trial type GORTs. It also offers a straightforward, computational analog of the proactive inhibition, comparable to the ExG model for reactive inhibition. Given the advantages of estimating the entire distribution of proactive inhibition over former point estimations, the researchers recommend considering the ALG model as the latest optimal choice to describe the distribution of proactive inhibition. \par  

\hspace{1 cm}\\
\textbf{Author Contributions:} The authors contributed to the study in the following manner: Conceptualization, M.S.; methodology, M.S.; formal analysis, M.S.; validation, M.E, M.S., A.D and R.S.; investigation, M.E., M.S., A.D and R.S.; data curation, R.S.; writing—original draft preparation, M.S.; writing—review and editing, M.S.,M.E., A.D., A.C., and R.S.; All authors have read and agreed to the submitted version of the manuscript.

%\section*{Acknowledgments}
%something
\section*{Disclosure Statement}
Mohsen Soltanifar, Michael Escobar, Annie Dupuis and Andre Chevrier have no financial interests to disclose.  Russell Schachar has equity in ehave and has been the on the Scientific Advisory Board(SAB) in Lilly and Highland Therapautic Inc, Toronto, Canada.
\section*{Funding}
This work has no external funding.

\section*{ORCID}

Mohsen Soltanifar:\ \ https://orcid.org/0000-0002-5989-0082\newline
Michael Escobar: \ \ \   https://orcid.org/0000-0001-9055-4709\newline
Annie Dupuis:\ \  \ \ \ \ \         https://orcid.org/0000-0002-8704-078X\newline
Andre Chevrier: \ \ \ \ https://orcid.org/0000-0002-4298-9529 \newline
Russell Schachar:\ \ \   https://orcid.org/0000-0002-2015-4395\newline

\section*{Abbreviations}
$
\begin{array}{ll}
\text{ADHD} & \text{Attention Deficit Hyperactivity Disorder} \\
\text{AL} & \text{Asymmetric Laplace distribution} \\
\text{ALG} & \text{Asymmetric Laplace Gaussian distribution} \\
\text{BEESTS} & \text{Bayesian Ex-Gaussian Estimation of Stop Signal RT distributions}\\
\text{BPA}& \text{Bayesian Parametric Approach}\\
\text{CDF}& \text{Cumulative Density Function}\\
\text{EM} & \text{Expectation Maximization}\\
\text{ExG} & \text{Ex-Gaussian distribution} \\
\text{GORT} & \text{Reaction Time in a go trial}\\
\text{GORTA} & \text{Reaction Time in a type A go trial}\\
\text{GORTB} & \text{Reaction Time in a type B go trial}\\
\text{HBPA} & \text{Hierarchical Bayesian Parametric Approach}\\
\text{IBPA} & \text{Individual Bayesian Parametric Approach}\\
\text{MGF} & \text{Moment Generating Function}\\
\text{ML} & \text{Maximum Likelihood}\\
\text{NL} & \text{Normal-Laplace distribution} \\
\text{OCD} & \text{Obsessive Compulsive Disorder}\\
\text{PDF} & \text{Prpobability Density Function}\\
\text{SE} & \text{Sequential Effect}\\
\text{SSD} & \text{Stop Signal Delay}\\
\text{SRRT} & \text{Reaction Times in a failed stop trial}\\
\text{SSRT} & \text{Stop Signal Reaction Times in a stop trial}\\
\text{SSRTA} & \text{Stop Signal Reaction Times in a type A stop trial}\\
\text{SSRTB} & \text{Stop Signal Reaction Times in a type B stop trial}\\
\text{SST} & \text{Stop Signal Task }\\
\text{$=^d$}& \text{Equality in distribution}\\
\text{$\oplus$} & \text{Sum of independent random variables}\\
\text{$\ominus$} & \text{Difference of independent random variables}
\end{array}
$

%%%%%%%%%%%%%%%%%%%%%%%%%%%%%%%%%%%%%%%%%%%%%%%%%%%%%%%%%55

\clearpage

\appendix

\textbf{Appendices}
\section{Proofs}
This appendix presents proofs for the new results presented in Section 3.1.
\subsection{Proof of Theorem 3.1.}
\textbf{Proof.}  Method(1): Let $\Delta GORT=GORT_B-GORT_A$ where $GORT_B,GORT_A$ are independent with ExG distribution with associated parameters $\theta_B=(\mu_B,\sigma_B,\tau_B ),\theta_A=(\mu_A,\sigma_A,\tau_A ),$ respectively. Then, by Definition \ref{def2.1}, Definition\ref{def2.2} and Theorem\ref{thm2.4} it follows that:
\begin{eqnarray}
m_{\Delta GORT} (t)	&=& m_{GORT_B} (t)\times m_{GORT_A} (-t)  \nonumber\\
                    &=& m_{ExG(\theta_B)} (t) \times m_{ExG(\theta_A)} (-t) \nonumber\\
                    &=& ((1-t\tau_B)^{-1} exp(\mu_B.t+\frac{\sigma_B^2}{2}t^2))\times ((1+t\tau_A)^{-1} exp(-\mu_A.t+\frac{\sigma_A^2}{2}t^2)) \nonumber\\
                    &=& ((1-t\tau_B)(1+t\tau_A) )^{-1} exp((\mu_B-\mu_A).t+\frac{\sigma_B^2+\sigma_A^2}{2}t^2) \nonumber\\
                    &=&  (1+(\tau_A-\tau_B)t-\tau_A\tau_B.t^2 )^{-1} exp((\mu_B-\mu_A).t+\frac{\sigma_B^2+\sigma_A^2}{2}t^2) \nonumber\\
                    &=& m_{AL(\tau_A,\tau_B)} (t)\times m_{N(\mu_B-\mu_A,\sigma_B^2+\sigma_A^2)} (t): \ \              -\tau_A^{-1}<t<\tau_B^{-1}.\nonumber
\end{eqnarray}
Accordingly, by Theorem\ref{thm2.5} it follows that:
$$\Delta GORT =^{d} AL(\tau_A,\tau_B) \oplus N(\mu_B-\mu_A,\sigma_B^2+\sigma_A^2)=^{d}ALG(\tau_A,\tau_B,\mu_B-\mu_A,(\sigma_B^2+\sigma_A^2)^{1/2}).$$
Method(2): Using Definition\ref{def2.1}, Definition\ref{def2.2}, Definition\ref{def2.3} and equation(\ref{eq12}) it follows that:
\begin{eqnarray}
\Delta GORT &=^{d}& GORT_B\ominus GORT_A \nonumber\\
            &=^{d}& ExG(\mu_B,\sigma_B,\tau_B)\ominus ExG(\mu_A,\sigma_A,\tau_A)\nonumber\\
            &=^{d}& (N(\mu_B,\sigma_B^2)\oplus Exp(\tau_B))\ominus (N(\mu_A,\sigma_A^2)\oplus Exp(\tau_A)) \nonumber\\
            &=^{d}& (N(\mu_B,\sigma_B^2)\ominus N(\mu_A,\sigma_A^2))\oplus (Exp(\tau_B)\ominus Exp(\tau_A)\nonumber\\
            &=^{d}& N(\mu_B-\mu_A,\sigma_B^2+\sigma_A^2)\oplus AL(\tau_A,\tau_B)\nonumber\\
            &=^{d}& AL(\tau_A,\tau_B)\oplus N(\mu_B-\mu_A,\sigma_B^2+\sigma_A^2) \nonumber\\
            &=^{d}& ALG(\tau_A,\tau_B,\mu_B-\mu_A,(\sigma_B^2+\sigma_A^2)^{1/2}).\nonumber
\end{eqnarray}
$\Box$

\subsection{Proof of Theorem 3.2.}
\textbf{Proof.}  By Definition\ref{def2.2} and Theorem\ref{thm2.4} for the nth cumulant of the ALG distribution with four parameters $(\alpha_1,\alpha_2,\mu,\sigma)$ we have:
\begin{eqnarray}
\kappa_n^{ALG(\alpha_1,\alpha_2,\mu,\sigma)}&=&	\kappa_n^{AL(\alpha_1,\alpha_2)}+\kappa_n^{N(\mu,\sigma^2)}\nonumber\\
&=&	((n-1)!((-\alpha_1 )^n+\alpha_2^n ))+(1_{(n=1)} (n)\mu+1_{(n=2)} (n) \sigma^2 ):        \ \ 1\leq n.\nonumber
\end{eqnarray}
Consequently, the descriptive and the shape statistics for the ALG distribution with four parameters $(\alpha_1,\alpha_2,\mu,\sigma)$, it follows that:

\begin{eqnarray}
Mean && E(ALG)=  \alpha_2-\alpha_1+\mu, \nonumber\\
Variance && Var(ALG)= \alpha_1^2+\alpha_2^2+\sigma^2,\nonumber\\
Skewness && \gamma_{ALG}=\frac{2(\alpha_2^3-\alpha_1^3 )}{(\alpha_1^2+\alpha_2^2+\sigma^2 )^{3/2}} , \nonumber\\
Kurtosis && \kappa_{ALG}=\frac{6(\alpha_1^4+\alpha_2^4 )}{(\alpha_1^2+\alpha_2^2+\sigma^2 )^2}\nonumber . 
\end{eqnarray}
Finally, the assertion follows for  $\alpha_1=\tau_A,\alpha_2=\tau_B,\mu=\mu_B-\mu_A,$ and  $\sigma^2=\sigma_B^2+\sigma_A^2.$ \newline
$\Box$

\subsection{Proof of Theorem 3.3.}

\textbf{Proof.} Given a four parametric $ALG(\alpha_1,\alpha_2,\mu,\sigma)$ for the proactive inhibition. Then, it can be written in the form:
$$ALG(\alpha_1,\alpha_2,\mu,\sigma)=\mu \oplus \sigma N(0,1) \oplus \alpha_2 Exp_2 (1)\ominus \alpha_1 Exp_1 (1)$$
where the random variables $N(0,1),Exp_i (1)\sim Exp(1)(i=1,2)$ are mutually independent. Accordingly, there are uncountably many solutions $(\mu_A,\sigma_A,\tau_A,\mu_B,\sigma_B,\tau_B)$ for the following equations:
$$\tau_A=\alpha_1,    \tau_B=\alpha_2,   \mu_B-\mu_A=\mu,    \sigma_B^2+\sigma_A^2=\sigma^2.$$
$\Box$

\subsection{Proof of Theorem 3.4.}

\textbf{Proof.} These are straightforward from the probability density function and \cite{M4}. \newline
$\Box$

\subsection{Proof of Theorem 3.5.}

\textbf{Proof.} Let, $\Delta GORT=^{d} ALG(\alpha_1,\alpha_2,\mu,\sigma) =^{d} AL(\alpha_1,\alpha_2 )\oplus N(\mu,\sigma^2 )$ where $(\alpha_1,\alpha_2,\mu,\sigma^2 )=(\tau_A,\tau_B,\mu_B-\mu_A,\sigma_A^2+\sigma_B^2 ).$ For, $X=^{d} AL(\alpha_1,\alpha_2 ),Y=^{d} N(\mu,\sigma^2 ),$ and two applications of Theorem\ref{thm2.6} for the following convex functions show that both components of the ALG distribution have increasing hazard functions:

\begin{eqnarray} 
-ln(f_X(t))&=& ln(\alpha_1+\alpha_2)\times (\frac{-t}{\alpha_1}) 1_{(-\infty,0]} + \frac{t}{\alpha_2} 1_{[0,\infty)}(t))\ \ -\infty<t<\infty,\nonumber\\
-ln(f_Y(t))&=& ln(\sqrt{2\pi}\sigma) + \frac{(t-\mu)^2}{2\sigma^2}.\ \ -\infty<t<\infty\nonumber
\end{eqnarray}
Accordingly, by an application of the Theorem\ref{thm2.7}, the plausible result follows.\newline
$\Box$

\section{ExG models parameters}
This appendix presents IBPA parameter estimations for cluster type ExG model.

\begin{table}[H]
\centering % centering table
\caption{ Mean posterior Ex-Gaussian parameters estimations across trial types by IBPA (n = 44). }  
\label{tableB1}
\resizebox{\textwidth}{!}{
\begin{threeparttable}[b]
\begin{tabular}{l ccccccccccc}  
\hline  
 & $\mu-$ &parameter && & $\sigma-$ &parameter & & & $\tau-$ & parameter &    \\
\hline 
$\#$ & $\mu_S$ & $\mu_A$ & $\mu_B$ & &$\sigma_S$ & $\sigma_A$ & $\sigma_B$ & & $\tau_S$ & $\tau_A$ & $\tau_B$   \\
\hline 
1&	357&	350&	372&&	32&	    35&	    14&&	86&	96&	68\\
2&	637&    599&	732&&	175&	170&	132&&   47& 48& 68\\
3&	469&	484&	411&&	60&	     57&	42&&	76&	73&	111\\
4&	597&	567&	631&&	163&	165&	96&&	66&	69&	149\\
5&	640&	618&	608&&	156&	133&	58&&	47&	62&	121\\
6&	452&	431&	469&&	108&	106&	64&&	66&	65&	135\\
7&	689&	668&	664&&	136&	130&	146&&   47& 60& 115\\
8&	665&	609&	660&&	145&	91&	    237&&   51& 103& 120\\
9&	543&	484&	640&&	166&	151&	118&&	120&	147&	157\\
10&	470&	468&	483&&	56&	    59&	    52&&	     98&	87&	156\\
11&	414&	399&	597&&	46&	    37&	    118&&	177&	168&	80\\
12&	557&	534&	597&&	132&	128&	146&&	53&	    58&	123\\
13&	550&	538&	564&&	137&	133&	98&&	    55&	     38&	190\\
14&	319&	318&	365&&	307&	295&    370&&	170&	137&	264\\
15&	421&	416&	437&&	61&	    56&	     90&&	138&	142&	149\\
16&	358&	342&	389&&	61&	    57&	    61&&	    48&	    56&	57\\
17&	594&	599&	561&&	130&	130&	133&&	78&	    62&	196\\
18&	467&	397&	747&&	229&	190&	299&&	127&	131&	159\\
19&	426&	426&	424&&	67&	    75&	    50&&	    102&	103&	110\\
20&	423&	449&	504&&	62&	    74&	    129&&	122&	65&	169\\
21&	521&	519&	487&&	144&	157&	96&&	    91&	    97&	125\\
22&	397&	346&	463&&	87&	    58&	    110&&	94&	    132&	101\\
23&	540&	525&	588&&	80&     78&	    79&&	    94&	    88&	128\\
24&	592&	571&	529&&	176&	136&	304&&	46&	    69&	180\\
25&	577&	459&	602&&	165&	70&	    244&&	69&	    181&	124\\
26&	562&	555&	694&&	79&	     75&	160&&	172&	154&	148\\
27&	446&	436&	541&&	71&	     60&	166&&	240&	236&	233\\
28&	486&	476&	629&&	82&	    64&	    196&&	172&	155&	151\\
29&	414&	363&	391&&	133&	66&	    213&&	62&	    115&	111\\
30&	486&	484&	541&&	87&	    86&	    146&&	141&	127&	181\\
31&	546&	502&	656&&	137&	118&	157&&	90&	    100&	137\\
32&	436&	421&	462&&	107&	109&	90&&	    72&	    81&	88\\
33&	452&	454&	458&&	38&	    46&	    40&&	    156&	156&	165\\
34&	404&	422&	408&&	105&	109&	42&&	    95&	    72&	92\\
35&	470&	549&	595&&	230&	200&	298&&	207&	171&	136\\
36&	429&	400&	448&&	116&	139&	95&&	    158&	163&	245\\
37&	521&	497&	507&&	89&	    130&	68&&	    112&	125&	222\\
38&	284&	271&	321&&	40&	    37&	    53&&	    100&	108&	91\\
39&	424&	432&	416&&	57&	    55&	    87&&	    70&	    52&	131\\
40&	419&	418&	476&&	52&	    53&	    196&&	145&	148&	105\\
41&	533&	537&	517&&	105&	151&	93&&	    72&	    35&	159\\
42&	388&	445&	497&&	53&	    145&	206&&	145&	57&	116\\
43&	506&	467&	539&&	97&	    82&	    100&&	66&	    96&	78\\
44&	842&	824&	822&&	175&	341&	165&&	34&	    95&	320\\
\hline
\end{tabular}
\begin{tablenotes}
     \item \footnotesize{Notes: $\mu_{S},\sigma_{S},\tau_{S}:$ ExG GORT parameters  for single cluster SST data; $\mu_{A},\sigma_{A},\tau_{A}:$ ExG GORT parameters  for type-A cluster SST data; $\mu_{B},\sigma_{B},\tau_{B}:$ ExG GORT parameters  for type-B cluster SST data;   IBPA: $\#$Chains = 3; Simulations = 20,000; Burn-in = 5,000 (for all parameters). }  
 \end{tablenotes}
\end{threeparttable}}
\end{table}

%%%%%%%%%%%%%%%%%%%%%%%%%%%%%%%%%%%%%%%%%%%%%%%%%%%%%%%%%%%%%%%%%% 

%%%%%%%%%%%%%%%%%%%%%%%%%%%%%%%%%%%%%%%%%%%%%%%%%%%%%%%%%%%%%%%%%%%%%%

%%%%%%%%%%%%%%%%%%%%%%%%%%%%%%%%%%%%%%%%%%%%%%%%%%%%%%%%%%%%%%%%%%%%%%%%%%%

\end{document}